\documentstyle[prb,aps,preprint]{revtex}
\author{Steffen Trimper}
\address{Fachbereich Physik\\
Martin-Luther-Universit\"at\\D-06099 Halle Germany}
\title{Kinetic induced phase transition}
\begin{document}
\draft
\date{\today}
\maketitle
\begin{abstract}
\noindent An Ising model with local Glauber dynamics is studied under the 
influence of additional kinetic restrictions for the spin-flip rates  
depending on the orientation of neighboring spins. Even when the static  
interaction between the spins is completely eliminated and only an external 
field is taken into account the system offers a phase transition at a finite 
value of the applied field. The transition is realized due to a competition 
between the activation processes driven by the field and the dynamical rules 
for the spin-flips. The result is based on a master equation approach in a 
quantum formulation.\\*[3cm]

\end{abstract}
		
\section{Introduction}

The dynamical properties of different models as Ising model, 
Sherington--Kirkpatrick spin glasses, influence of time dependent 
external fields have been the subject of continuous interest \cite{priv}.  
Because the conventional Ising model is a static, equilibrium model, 
a dynamical generalization was first considered by Glauber \cite{gl} who 
introduced the single-spin-flip kinetic Ising model for describing relaxation 
towards equilibrium. The dynamics is generated by a coupling to a heat 
bath at a fixed temperature. Hence the rates of the dynamical processes are 
constrained to satisfy detailed balance. Based on that the static interaction 
energy of the spins is included in the flip-rates. In this way the equilibrium 
phase transition inherent for the Ising system in $d > 1$ is also manifested 
in the dynamical approach such as within the behavior of the relaxation time 
and the correlation function.\\
In the present paper we discuss the possibility that the kinetic Ising system 
undergoes a phase transition although the mutual static interaction between 
the spins will be neglected completely. The transition is purely mediated by 
the underlying kinetics. Special attention will be paid to dynamical 
restrictions for the spin-flip rates instead of including an interaction. 
As a consequence of such constraints combined with the orientation of 
the applied field (activation energy) there occurs a competing situation 
leading to the aforementioned kinetic induced phase transition.\\ 
To be specific let us consider a spin system in an external time independent 
field which is subject to a dynamical process (spin-flip process). However, 
it is assumed that the flip rates does not depend on the neighboring spin 
configuration via the static energy between adjacent spins manifested by a 
Hamiltonian. Alternatively, kinetic restrictions are included 
independently on their energy. The environment of a fixed spin is oriented in a 
self-adaptive manner only by the kind of kinetic processes. Formally, this can be achieved 
by chosing the flip rates depending on the state of the neighboring spins. 
Furthermore, the flip rates are also determinated by the temperature and the 
activation field. To formulate the master equation under the influence of 
constraints we use a representation 
of the underlying kinetic equation in terms of quantum Pauli--operators. 
Based on the analogy to quantum mechanics the problem can be studied in a 
compact form \cite{doi}, for a recent review see \cite{sti,mg}. The method 
allows the inclusion of the above mentioned restrictive conditions in a 
transparent manner which had been demonstrated for the so called Fredrickson 
Andersen model \cite{fa,fb}, see \cite{st3}.

\section{The model}

Let us use a lattice gas description with the occupation number 
at a certain lattice site defined by $n_i = 0,1$. Due to the relation to 
the spin variable, $S_i = 1 - 2 n_i$, an empty site corresponds to an up 
orientation whereas an occupied site is related to spin-down. The local flip 
dynamics can be characterized by an evolution operator written in terms of 
Pauli--operators \cite{sctr} 
\begin{equation}
L_i = \lambda (1 - d^{\dagger}_i)d_i + \gamma (1 - d_i)d^{\dagger}_i 
\label{fli1}
\end{equation}
The first term gives a nonzero contribution in case of a flip from a spin-down 
to a spin-up state (or from an occupied to an empty site) with the rate 
$\lambda$ whereas the second term including the rate $\gamma$ 
is responsible for the inverse process, i.e. the flip from spin-up to 
spin-down states. Obviously, the rates $\lambda$ and $\gamma$ can depend  
on the temperature of a heat bath coupled to the system, on the 
mutual interaction between the spin degrees of freedom, as well as on the 
neighboring spin configuration in a kinetic manner discussed below.  
The coupling to a heat bath is easily realized by the following 
modification of the evolution operator \cite{sctr}
\begin{equation}
L = \nu \sum \left[ (1 - d_i) \exp(-\beta H/2) d^{\dagger}_i 
\exp(\beta H/2)  + (1 - d^{\dagger}_i) \exp(-\beta H/2) d_i \exp(\beta H/2) \right] 
\label{fli2}
\end{equation}
The procedure is similar to a 'thermalized' Heisenberg picture in quantum 
mechanics. As the simplest case let us consider a coupling to an activation 
field of the strength $\varepsilon$
\begin{equation}
H = \sum_i \varepsilon n_i
\label{field}
\end{equation}
If $\varepsilon > 0$ for each lattice site the spin-up orientation is 
preferred. In the opposite case $\varepsilon < 0$ the spin-down state is more 
favored. 
The algebraic properties of Pauli--operators can be used to calculate for 
instance $\exp(-\beta H/2) d^{\dagger}_i \exp(\beta H/2) = 
d^{\dagger}_i \exp(1/2 \beta \varepsilon)$. 
Comparing eq.(\ref{fli1}) and eq.(\ref{fli2}) the thermalization leads to 
temperature dependent activation rates which can be written in the case 
considered by eq.(\ref{field}) as
\begin{equation}
\lambda = \nu \exp(\varepsilon/2T) \quad \gamma = \nu \exp(-\varepsilon/2T).
\label{ra}
\label{fli3}
\end{equation}
It should be remarked that the transition rate from down to up state 
is realized at low temperatures with a larger probability than the 
opposite process (in case $\varepsilon > 0$). For high temperatures, 
$T \to \infty$, both flip-processes appear with the same probability. 
Before the model is extended by introducing kinetic restrictions let us 
demonstrate its applicability.\\ 
The evolution operator $L$ obeys the equation 
\begin{equation}
\partial_t \mid F(t)\rangle = L \mid F(t) \rangle
\label{fo1}
\end{equation}
where the state vector $\mid F(t) \rangle$ is related to the probability 
distribution function $P(\vec n,t)$ according to 
$P(\vec n,t) = \langle \vec n \mid F(t) \rangle$ with 
a complete set of basic-vectors $\mid \vec n \rangle$ composed of 
Pauli--operators. The probability distribution function is assumed to follow 
a master equation written in the symbolic form 
\begin{equation}
\partial_t P(\vec n,t) = \hat{L} P(\vec n,t).
\label{ma}
\end{equation}
The evolution operator $L$ in eq.(\ref{fo1}) can be mapped onto the operator 
$\hat{L}$ in the master equation. Such an approach had been introduced by 
Doi \cite{doi} using Bose operators \cite{pe}. A generalization to Pauli--operators had 
been proposed \cite{gra,gw,satr,sd,scsa,aldr}, for a recent review see 
\cite{sti,mg}.
Following \cite{doi,gra,pe,satr,aldr} the probability distribution 
$P(\vec n,t)$ is related to a state vector $\mid F(t) \rangle$ in a 
Fock-space according to $P(\vec n,t) = \langle \vec n\mid F(t)\rangle$ with 
the basis-vectors $\mid \vec n \rangle$ composed of second quantized operators. 
A further extension to an p--fold occupation number is 
possible and had been discussed in \cite{sctr2}.\\
\noindent The relation between the quantum--like formalism and the 
probabilistic approach is given by
\begin{equation}
\mid F(t) \rangle = \sum_{n_i} P(\vec n,t) \mid \vec n \rangle.
\label{fo2}
\end{equation}  
As it was shown firstly by Doi \cite{doi} the average of an arbitrary physical 
quantity $B(\vec n)$ is defined by the average of the corresponding operator 
$B(t)$
\begin{equation}
\langle B(t) \rangle = \sum_{n_i} P(\vec n,t) B(\vec n) = 
\langle s \mid B \mid F(t) \rangle 
\label{fo3}
\end{equation} 
with the state function $\langle s \mid = \sum \langle \vec n \mid$. Using the 
relation $\langle s \mid L = 0$ the evolution equation for an operator 
$A$ can be written  
\begin{equation}
\partial_t \langle B \rangle = \langle s \mid [B,L] \mid F(t) \rangle.
\label{kin}
\end{equation}
It should be stressed that all the dynamical equations covering the 
classical problem are determined by the commutation rules of the underlying 
operators and the structure of the evolution operator $L$. 
In our case the dynamics of the model is given by spin-flip processes 
indicating a change of the local occupation number or the spin orientation, 
respectively.\\ 
\noindent Based on the above introduced approach the evolution equation for the 
occupation number operator or the averaged 
spin for the simple flip process eq.(\ref{fli2}) is easily derived. We find
\begin{equation}
\nu^{-1} \partial \langle n \rangle = - \langle n \rangle e^{\varepsilon/2T} 
+ \langle 1 - n \rangle e^{-\varepsilon/2T}. 
\end{equation}
The solution in a spin representation is well known and reads
\begin{equation}
\langle S(t) \rangle = \tanh(\frac{2 \varepsilon}{T}) + S_0 \exp(-t/\tau)\quad 
\mbox{with}\quad \tau = \frac{1}{2 \cosh(\frac{2 \varepsilon}{T})}. 
\end{equation}
The first term represents the stationary state whereas the second one 
describes the exponential decay with a relaxation time $\tau$. For low 
temperatures one finds $\tau \propto \exp(-\frac{2 \varepsilon}{T})$. In the 
opposite limit the relaxation time shows an algebraic dependence on the 
activation energy 
$\tau \propto \frac{1}{2}( 1 - 2 (\frac{\varepsilon}{T})^2 )$. The results 
are independent on the sign of the activation energy.\\
Now we discuss an extension of the model introducing constraints within 
dynamical rules specified below. It will be demonstrated that in this case 
the sign of $\varepsilon$ leads to a complete different behavior of the 
system. To this aim an anisotropic model is studied where the kind of 
anisotropy is not originated by the static interaction. Instead of that the 
flip processes will be modified in the following manner. 
Whereas the flip process from spin-down to spin-up orientation should be 
possible whenever a double occupancy is avoided which is automatically 
guaranteed by the algebraic properties of the Pauli--operators, the inverse 
process, spin-up to spin-down flip, should be restricted. Such a flip is only 
allowed when the neighboring configuration fulfills the condition  
\begin{equation}
\frac{1}{2}\sum_{j(i)} (1 + S_j) \le f
\label{kr}
\end{equation}
A local spin flip at the lattice site $i$ is only permitted when the number of 
nearest neighbors denoted by $j(i)$ satisfies the last condition. Here $f$ is 
a certain but fixed number specified below. Physically the confinement 
means that flips from up- to down-spin orientations are reduced 
when the total number of adjacent spins with an spin-up alignment does not 
exceed the number $f$. Such kinetic restrictions are already introduced by 
Fredrickson and Andersen \cite{fa} within a f-spin facilitated kinetic  
Ising model which is a candidate describing the ultraslow dynamics in the 
vicinity of a the glass transition \cite{sr1,sr2,st2,st3}.\\
Here, we analyse an anisotropic version of the model by assuming that only one 
of the flip rates, $\lambda$ or $\gamma$ are modified which is different from 
that one discussed for glassy materials. 
The kinetic restriction 
manifested by eq.(\ref{kr}) means that a local change from a spin-up state 
to a spin-down state is not possible when more than f nearest neighbors 
with respect to a given site are in the up position. Mathematically, the 
confinement is realized by assuming that the corresponding transition rate, 
here $\gamma$, will be replaced by $\gamma \to \gamma n_{r_1} \dots n_{r_f}$. 
As the simplest choice let us chose $f= z/2$ where $z$ is the coordination 
number. This leads to the following evolution operator
\begin{eqnarray}
L &=& \nu \sum \chi_{ij_1 \cdots j_f} \left[(1 - d_i) \exp(-\beta H/2) 
d^{\dagger}_i \exp(\beta H/2)  n_{j_1} \cdots n_{j_f} \right] \nonumber\\ 
&+& \left[ (1 - d^{\dagger}_i) \exp(-\beta H/2) d_i \exp(\beta H/2) 
\right]
\label{fli4}
\end{eqnarray}
where $\chi_{ij_1 \cdots j_f}$ is only nonzero when all the indices indicates 
nearest neighbor sites. Here we consider the case $f=2$ which corresponds to 
a cubic lattice in two dimensions. The general results seem to be 
independent on the special realization. As the main feature of the model there 
occurs a conflicting situation between the static preferred orientation of 
the spins manifested in the sign of the activation energy $\varepsilon$ and 
the kinetic rules. That case will be discussed in detail in the forthcoming 
section.

\section{Anisotropic model}

Using the algebraic properties of Pauli--operators the evolution equation 
for the averaged occupation number reads
\begin{equation}
\partial_t \langle n_j \rangle = 
\nu \sum_{r,s} \chi_{jrs}[ \exp(-\varepsilon_r/2T) 
\langle(1-n_j) n_r n_s \rangle - \exp(\varepsilon_j/2T) \langle n_r \rangle] 
\label{a1}
\end{equation}
As already demonstrated in \cite{sctr} the mesoscopic dynamic equations  
within the classification introduced by Hohenberg and Halperin \cite{hh} 
is rederived in a coarse grained approximation. 
In the present case the equation is written in the form
\begin{equation}
\partial_t \langle n_j \rangle = - e^{\varepsilon/2T}z(z-1) \langle n_j \rangle + e^{-\varepsilon/2T} 
\sum_{rs}\chi_{jrs} [ \langle n_r \rangle \langle n_s \rangle -  \langle n_j \rangle \langle n_r \rangle 
\langle n_s \rangle ]
\label{a1a}
\end{equation}
It consists of a local part (linear term) and nonlocal terms. 
After Fourier transformation and introducing the abbreviation 
$\langle n(\vec k, t) \rangle \equiv n_k(t)$ 
the last equation reads
\begin{eqnarray}
\nu^{-1}\partial n_k(t) &=& - e^{\varepsilon/T} n_k(t) z(z-1)  \nonumber\\
&+&e^{-\varepsilon/T} 
\sum_{p,q} \left[ - \epsilon _{pq} n_{k-p-q}(t) n_p(t) n_q(t) + \epsilon_{k-pp} n_{k-p}(t) n_p(t) \right]
\label{a1b}
\end{eqnarray}
Here we have introduced the dispersion relation ($l$ is the lattice constant) 
\begin{equation}
\epsilon_{pq} = 4[\cos p_xl \cos q_yl + \cos p_yl \cos q_xl] + 2 [\cos(p_x - q_x)l + \cos(p_y - q_y)l ] 
\label{dis}
\end{equation}
The homogeneous stationary state $n_0$ obeys the equation
\begin{equation}
0 = -\lambda n_0 + \gamma (n^{2}_0 - n^{3}_0 )
\label{a2}
\end{equation}
This equation yields three stationary points 
\begin{equation}
n_1 = 0 \quad n_{2/3} = \frac{1}{2}[1 \pm \sqrt{1 - 4 \rho}] 
\quad \mbox{with} \quad \rho = \frac{\lambda}{\gamma} = \exp(\varepsilon/T). 
\end{equation}
Making the conventional ansatz $n(k,t) = n_0 \delta(\vec k) + m(\vec k,t)$ we get 
\begin{eqnarray}
\partial_t m(\vec k ,t) &=& - \Lambda(\vec k) m(\vec k, t) \quad \mbox{with} \nonumber\\  
\Lambda(\vec k,t) &=& \lambda \epsilon_{00} + \gamma [ (2 \epsilon_{k0} + \epsilon_{00}) n_0^2 - 
2 \epsilon_{k0} n_0 ]
\label{stab}
\end{eqnarray}
From the linear stability analysis we conclude that the spin-up orientation 
$n_1 = 0$ is always stable independently on the sign of the activation energy 
$\varepsilon$. For the nonzero solutions the function $\Lambda(\vec k)$ 
introduced in eq.(\ref{stab}) can be written in the form
\begin{eqnarray} 
\Lambda(\vec k) &=& \Lambda(0) +\lambda \epsilon_{00}( 2 - \cos(k_xl) + \cos(k_yl) ) \nonumber\\
\mbox{with}\quad \Lambda(0) &=& \epsilon_{00} \gamma(n_0 - 2\rho)
\label{stab1}
\end{eqnarray}
Obviously, the sign of gap $\Lambda(0)$ in eq.(\ref{stab1}) is responsible for the stability.  
Whenever $\varepsilon > 0$ the both nonzero solutions $n_2$ and $n_3$ are 
unstable. Hence, we find that in this kind of an anisotropic model the up 
orientated spin state is the single stable one. The result is expected 
because the static field (activation energy) favors the spin-up state, 
$\varepsilon > 0$, and the dynamic rules support the same orientation with a 
larger probability as the opposite one. 
The result is independently on the temperature and the strength of the 
activation energy.\\
The situation is complete different analysing the case $\varepsilon < 0$. 
The static energy would favor the down orientation of the spins due to 
the sign of the field. In contrast to that the dynamical rules tend to avoid 
such a state because states with a down-spin orientation are reduced 
by the kinetic restrictions. As the result of the competition between both 
processes there appears a critical temperature 
$T_c = \frac{|\varepsilon|}{\ln 4}$. Above this temperature only the 
state $n_1 = 0$ is a stable one. For $T < T_c$ both states $n_1$ and $n_2$ 
are stable, where $n_2$ can be rewritten 
\begin{equation}
n_2 = \frac{1}{2} \left[ 1 + \sqrt{1 - \exp(-\mid \varepsilon \mid (T^{-1} - T_c^{-1})} \right]
\end{equation}
The situation described in the case $\varepsilon < 0$ is remarkable. For high 
temperature the spin-up state is realized. Whereas the flip rates in this case 
are of the same order the static term would favor 
a spin-down state. Due to the kinetic restrictions such a state is not 
realized, or with other words the dynamic rules with the above introduced 
constraints work much stronger than the activation field. The phase transition 
is exclusively induced by the kinetics. For very low temperatures the 
flip-rate to an up state (promoted by $\lambda$) is very small compared with 
the rate for the opposite process. Insofar, the 
kinetics favor the occurrence of the spin-down orientation. From this point of 
view one should expect that for $T < T_c$ only the state $n_2 \ne 0$ is 
realized. This intuitive result can be confirmed by comparing the energy of 
both states. Starting from eq.(\ref{a1a}) we find an energy function $F$ which 
reads 
\begin{equation}
\frac{F}{z(z-1)} =  \frac{\lambda}{2} n^2 - \frac{\gamma}{3} n^3 + \frac{\gamma}{4} n^4
\end{equation}
where we have considered only the homogeneous case which seems to be 
sufficiently. The function $F$ plays the role of the free energy. However, it 
should be stressed that F is not simply a Ginzburg-Landau functional because 
different to that one the prefactor of the quadratic part does not 
change its sign at the phase transition temperature ($\lambda > 0$) reflecting 
the fact that the system does not possess a phase transition in equilibrium. 
Now let us calculate the energy difference $\Delta F = F(n_2) - F(n=0)$ with 
the result 
\begin{eqnarray}
\Delta F &<& 0 \quad \mbox{for}\quad T < T_0 = \frac{\mid \varepsilon\mid}{\ln(4.5)} \nonumber\\
\Delta F &>&0 \quad \mbox{for}\quad T > T_0
\end{eqnarray}
From here we conclude, that the solution $n_1 = 0$, spin-up orientation,
is energetically preferred within the interval $T_0 < T < T_c$ whereas for 
$T < T_0$ the nonzero solution $n_2$ is stable and is related to the lowest 
energy. The situation is comparable with a first order phase transition, but 
as already stressed, the transition is originated by the kinetic restrictions 
and it is not driven by the energy.\\ 

\noindent The above introduced model can be easily extended by considering 
other restrictions. For instance the flip process spin-down to spin-up can be 
reduced when $f = 2$ neighbors are in the down position. In our approximation 
it means that the transition rate $\lambda$ is replaced by $\lambda (1 - n)^2$. 
The result does not change when simultaneously one substitutes $\lambda$ by 
$\gamma$ and $\varepsilon$ by $-\varepsilon$. Another extension 
can be realized when both flip processes are confined for instance by 
assuming $\lambda \to \lambda (1 - n)^2$ and $\gamma \to \gamma n^2$. When 
the activation energy is positive, $\varepsilon > 0$, one finds three stable 
equilibrium solutions $n_1 = 0, n_2 = 1, n_3 = \frac{\lambda}{\lambda + \gamma}$ 
with $F(n_1) < F(n_2) < F_(n_3)$. In the opposite case of a negative activation 
field $\varepsilon < 0$ we get two stable solutions $n_1 = 0$ and $n_2 = 1$ with 
$F(n_2) < F(n_1)$. A third solution $n_3 = \frac{\lambda}{\lambda + \gamma}$ is 
only stable for $T > T_c$. Moreover, other realization for constraints 
does not lead to significant changes.

\section{Conclusions}

In the present paper we have considered a system with competing forms 
of interactions. Whereas the static energy is simply the energy for spins in a field which does 
not lead to an equilibrium phase transition the dynamical rules for the thermal 
activated flip processes are supplemented by additional restrictions. These 
restrictions are self organized by the orientation of neighboring spins with 
respect to a given one. 
Whenever the field and the kinematics favor the same spin orientation a phase 
transition is suppressed. In the opposite case, when orientation preferred 
by the static activation field will be restricted or even when a certain order 
is prevented by the dynamics the system offers a phase 
transition. There appear two stable solutions below a critical temperature 
$T_c$ where this temperature is purely determined by the field and not by 
quantities included in an energy functional of Ginzburg-Landau type. To be 
specific, we have analysed the case that a down-orientation 
(occupation number $n = 0$) is supported by the orientation of the 
field $\varepsilon$. The flip-rate $\lambda$ responsibly for a spin-flip from 
down-to up-orientation is purely given by the activation process. However the 
inverse process, a spin-flip from up- to down-orientation 
(rate $\gamma$) is additional constrained by the dynamical alignment of 
neighboring spins. In particular, the mentioned flip process is strongly 
restricted when more than a half of the neighboring 
spins of a fixed one is already in an up-position. This conflicting situation 
leads to the phase transition at $T_c$. 
For low temperatures the relation $\gamma \gg \lambda$ is valid. The flip 
process, spin-up to spin-down, is realized with a much higher rate as the 
inverse process. Because the restrictions favors an up-orientation we find 
for a finite temperature that the state $n \simeq 1$ is the stable one. This 
state is always realized for $T < T_0$. However, there exists a second stable 
solution with spin-up orientation but with a higher energy. On that manner 
this state is only a metastable one. At $T_0$ both states are simultaneously 
stable and coexist with the same energy. For $T > T_0$ the up-orientation is 
favored with a lower energy. When $T > T_c$ only the up-orientation is stable, 
that means the restrictive dynamics determines the behavior of the model.\\
It seems to be for the first time that such a kinetic mediated phase 
transition is discussed. As a further step we should take into account 
fluctuations to observe the detailed behavior in the vicinity of the branching 
point.

\end{document}